\documentclass[prl,aps,twocolumn,showpacs,superscriptaddress]{revtex4-1}
\usepackage{amsmath}
\usepackage{graphicx}
\usepackage{dcolumn}
\usepackage{bm}
\usepackage{float}
\usepackage[colorlinks,
            linkcolor=blue,
            anchorcolor=blue,
            citecolor=blue
            ]{hyperref}
\begin{document}


\title{Phagraphene: A Low-energy Graphene Allotrope composed of 5-6-7 Carbon Rings with Distorted Dirac Cones}

\author{Zhenhai Wang}
\email{wangzh@njupt.edu.cn (Z.W.)}
\affiliation{Peter Gr\"{u}nberg Research Center, Nanjing University of Posts and Telecommunications, Nanjing 210003, China}
\affiliation{Department of Geosciences, Center for Materials by Design, and Institute for Advanced Computational Science, Stony Brook University, Stony Brook, NY 11794, USA}


\author{Xiang-Feng Zhou}
\affiliation{Department of Geosciences, Center for Materials by Design, and Institute for Advanced Computational Science, Stony Brook University, Stony Brook, NY 11794, USA}
\affiliation{School of physics and Key Laboratory of Weak-Light Nonlinear Photonics, Nankai University, Tianjin 300071, China}

\author{Xiaoming Zhang}
\affiliation{School of Physics and State Key Laboratory of Crystal Materials, Shandong University, Jinan 250100, China}

\author{Qiang Zhu}
\affiliation{Department of Geosciences, Center for Materials by Design, and Institute for Advanced Computational Science, Stony Brook University, Stony Brook, NY 11794, USA}

\author{Huafeng Dong}
\affiliation{Department of Geosciences, Center for Materials by Design, and Institute for Advanced Computational Science, Stony Brook University, Stony Brook, NY 11794, USA}

\author{Mingwen Zhao}
\email{zmw@sdu.edu.cn (M.Z.)}
\affiliation{School of Physics and State Key Laboratory of Crystal Materials, Shandong University, Jinan 250100, China}

\author{Artem R. Oganov}
\email{artem.oganov@stonybrook.edu (A.R.O.)}
\affiliation{Department of Geosciences, Center for Materials by Design, and Institute for Advanced Computational Science, Stony Brook University, Stony Brook, NY 11794, USA}
\affiliation{Skolkovo Institute of Science and Technology, Skolkovo Innovation Center, Bldg.3, Moscow 143026, Russia}
\affiliation{Moscow Institute of Physics and Technology, 9 Institutskiy Lane, Dolgoprudny City, Moscow Region 141700, Russia}
\affiliation{School of Materials Science, Northwestern Polytechnical University, Xi'an 710072, China}


\begin{abstract}
Using systematic evolutionary structure searching we propose a new carbon allotrope, phagraphene[\textbf{f{\ae}'gr{\ae}fi:n}], standing for \textbf{p}enta-\textbf{h}exa-hept\textbf{a}-graphene, because the structure is composed of 5-6-7 carbon rings. This two-dimensional (2D) carbon structure is lower in energy than most of the predicted 2D carbon allotropes due to its \emph{sp}$^2$-hybridization and density of atomic packing comparable to graphene. More interestingly, the electronic structure of phagraphene has distorted Dirac cones. The direction-dependent cones are further proved to be robust against external strain with tunable Fermi velocities.
\end{abstract}

\maketitle


Graphene, as the most stable two dimensional (2D) form of carbon, exhibits a number of unusual electronic and spintronic properties \cite{R01}, such as high carrier mobility \cite{R02} and quantum Hall effects \cite{R03,R04}. Its honeycomb lattice, which consists of two equivalent carbon sublattices with hexagonal symmetry, plays a crucial role in the formation of the Dirac cones with linear dispersion.
Successful preparation of graphene in 2004 \cite{R05} has inspired further searches for other 2D Dirac materials. Among the predicted 2D Dirac materials \cite{R06}, are silicene \cite{R07}, germanene \cite{R07}, silicon germanide monolayer \cite{R08}, graphynes \cite{R09,R10}, \emph{Pmmn} boron \cite{R11}, \emph{m}-TiB$_2$ \cite{R12}, so-MoS$_2$ \cite{R13}, \emph{etc.}, while only Dirac cones in graphene have been actually confirmed experimentally \cite{R14}. The stable and robust 2D carbon backbones motivate the searches for other Dirac carbon allotropes.

Graphynes, composed of \emph{sp}- and \emph{sp}$^{2}$-hybridized atoms, were first proposed in 1987 \cite{R15} with various structures and high thermal stability. These graphynes can be metalic, semimetalic or semiconducting \cite{R16,R17}. Up to 2012, Malko \emph{et al.} \cite{R09} reported that  $\alpha$-, $\beta$- and 6,6,12-graphyne could have Dirac cones. Huang \emph{et al.} have tried to derive a criterion to explain the physical origin of the existence or absence of Dirac cones in graphynes, in which they successfully predicted two other graphynes with Dirac cones, 14,14,14-graphyne and 14,14,18-graphyne \cite{R18}. After that, $\delta$-graphyne was proposed as another Dirac carbon allotrope, which was demonstrated to be a carbon topological insulator superior to graphene \cite{R10}. For all these graphynes with Dirac cones, $\alpha$-, $\beta$- and  $\delta$-graphynes exhibit hexagonal symmetry, while 6,6,12-, 14,14,14-, and 14,14,18-graphynes have rectangular lattices, which suggests that hexagonal symmetry is not a necessary precondition for the presence of Dirac cones.

However, the \emph{sp}-hybridization of carbon atoms involved in these graphynes is energetically disadvantageous compared with that of \emph{sp}$^{2}$- and \emph{sp}$^{3}$-carbon atoms. Several carbon allotropes composed of \emph{sp}$^{2}$- and \emph{sp}$^{3}$-carbon atoms have also been predicted to have Dirac cones. Liu \emph{et al.} proposed ``T-graphene", named after a buckled carbon sheet with tetrarings, as a 2D Dirac material \cite{R19}. Although it was demonstrated to be metallic in a subsequent work \cite{R20}, this opens a possibility to find novel 2D Dirac graphene allotropes with multi-member rings. In 2014, Xu \emph{et al.} designed three rectangular allotropes by re-constructing graphene, named as S-, D-, and E-graphene. These \emph{sp}$^{2}$- and \emph{sp}$^{3}$-carbon networks were predicted to be stable and have Dirac cones \cite{R21}. They also demonstrated that the Dirac cone and the carrier linear dispersion are common features of these 2D carbon allotropes.

In this letter, we predict a new carbon allotrope, phagraphene, which is composed of 5-6-7 carbon rings. This planar carbon allotrope is energetically comparable to graphene and more favorable than other carbon allotropes proposed in previous works, due to its \emph{sp}$^{2}$-hybridization and dense atomic packing. Both density functional theory (DFT) and tight binding (TB) model confirm the distorted Dirac cone in the first Brillioun zone (BZ) for this 2D carbon structure. The direction-dependent Dirac cone is further proved to be robust against external strain with tunable Fermi velocities.

Systematic 2D structure searches were performed using the \emph{ab initio} evolutionary algorithm USPEX \cite{R11,R22,R23,R24} with 6, 8, 10, 12, 14, 16, 18, 20, 22 and 24 carbon atoms per unit cell. The initial thickness was set to zero, since \emph{sp}- and \emph{sp}$^2$-carbon allotropes prefer planar structures at ambient pressure. The newly produced structures were all relaxed, and the energies were used for selecting structures as parents for the new generation of structures. Structure relaxations used the projector-augmented-wave (PAW) method \cite{R25}, as implemented in the Vienna \emph{ab initio} simulation package (VASP) \cite{R26,R27}. The exchange-correlation energy was treated within the generalized gradient approximation (GGA), using the functional of Perdew, Burke, and Ernzerhof (PBE) \cite{R28}.

For calculations of specific 2D carbon allotropes in VASP, a kinetic energy cutoff of 600 eV was adopted. BZ integrations were carried out using Monkhorst-Pack sampling grids with resolution of 2$\pi$$\times$0.04 {\AA}$^{-1}$ for structure optimizations. The atomic positions and lattice constants were optimized using the conjugate gradients (CG) scheme until the force components on each atom were less than 0.01 eV/{\AA}. An adequate \emph{k}-point sampling (30$\times$30$\times$1) was employed for charge density distributions and band structure calculations. Phonon calculations by supercell approach as implemented in the PHONON code \cite{R29} was used to examine dynamical stability, and first-principles molecular dynamics simulations under constant temperature and volume (NVT) were performed to check thermal stability.

\begin{figure}
\begin{center}
\includegraphics[width=0.825\columnwidth]{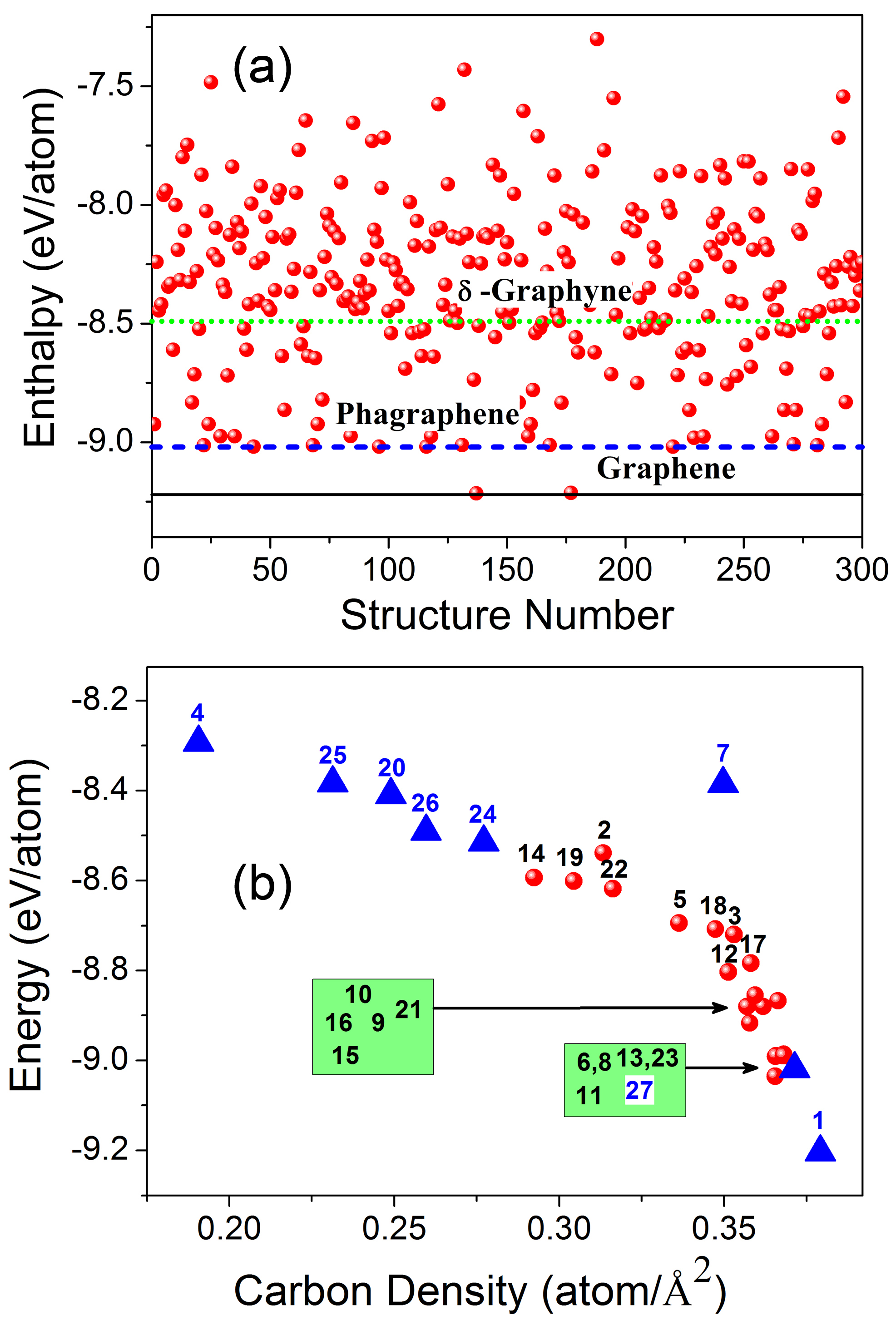}
\end{center}
\caption{\label{fig1}
(color online) (a) Typical enthalpy evolution for a 20-atom carbon system during an evolutionary structure search; (b) Energies and planar atomic densities for most of 2D carbon allotropes re-obtained from our evolutionary structure searches \cite{R30}. Allotropes with Dirac cones are labeled as blue triangles.}
\end{figure}

During the structure searches, graphene always occupies the lowest position in the enthalpy evolutions. We present a typical example (20 atoms/cell) in Fig. \ref{fig1}(a). Besides graphene, phagraphene and $\delta$-graphyne \cite{R10} with a 20-atom primitive cell were generated during this search. As we can see, the energies for graphene, phagraphene and  $\delta$-graphyne are -9.23, -9.03 and -8.49 eV/atom, respectively. With different atom numbers in cell, many 2D carbon structures with hybrid \emph{sp}- and \emph{sp}$^2$-carbon atoms were generated in our searches, but these structures usually have higher energies than \emph{sp}$^2$-carbon allotropes. For \emph{sp}$^2$-carbon structures, many of them are composed of penta-, hexa-, hepta-, and octa- (5-, 6-, 7-, and 8- ) carbon rings. Most of the planar carbon allotropes proposed in previous works can be reproduced in our systematic searches. Their energies and planar atomic densities are plotted in Fig. \ref{fig1}(b). It is noted that Dirac allotropes are mostly at much higher energies compared with graphene (denoted as 1), and structures with higher energies usually have lower planar atomic densities. The planar atomic density of phagraphene (denoted as 27) is 0.37 atoms/{\AA}$^2$, slightly smaller than 0.38 atoms/{\AA}$^2$ of graphene. Thanks to this dense packing of \emph{sp}$^2$-carbon atoms, it has lower energy than most of before predicted 2D carbon allotropes.

\begin{figure*}
\begin{center}
\includegraphics[width=1.5\columnwidth]{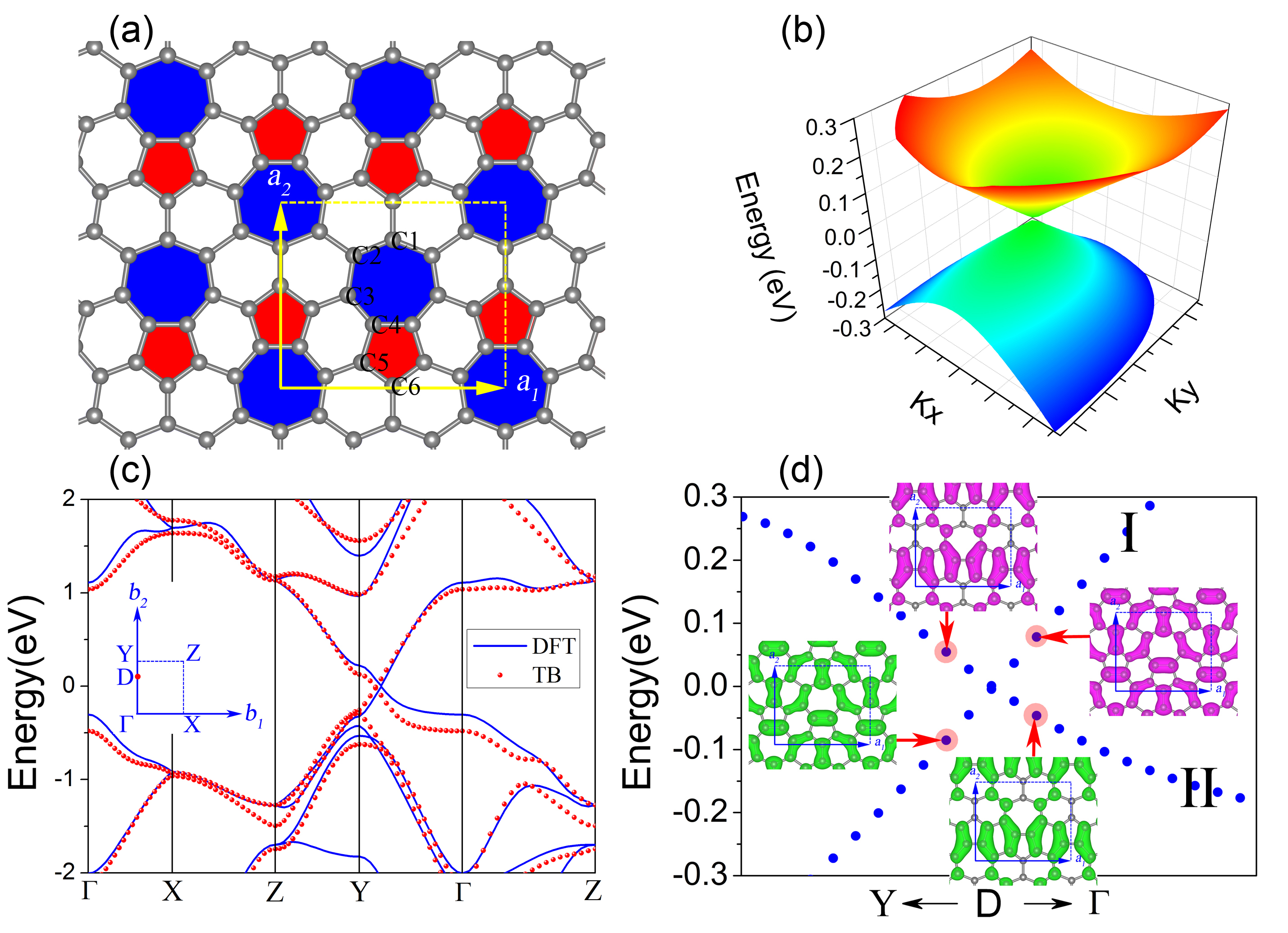}
\end{center}
\caption{\label{fig2}
(color online) (a) Structure of phagraphene with notable space-inversion symmetry, C1-C6 are inequivalent carbon atoms in its unit cell; (b) The distorted Dirac cone formed by the valence and conduction bands in the vicinity of the Dirac point; (c) The comparison of band structures from density functional theory (DFT) and tight binding (TB) model. Inserted first Brillouin zone (BZ) with high-symmetry \emph{k} points: $\Gamma$ (0,0,0), X (0.5,0,0), Z (0.5,0.5,0), Y (0,0,5,0) and Dirac point: D (0, 0.377,0); (d) Charge density distributions near the distorted Dirac cone, both Dirac bands (denoted as I and II) are from \emph{p$_z$} orbitals of \emph{sp}$^2$-carbon atoms. Fermi level has been set to zero.}
\end{figure*}

Figure \ref{fig2}(a) shows its primitive cell and the periodic structure with notable space-inversion symmetry. The structure of phagraphene is composed of 5-6-7 carbon rings. Unlike \emph{P6m} of graphene and $\delta$-graphyne, the plane group of phagraphene is \emph{Pmg}. Being surrounded by 6-carbon rings, the pairs with 5- and 7-carbon rings invert to each other and arrange in a rectangular sublattices. Fig. \ref{fig2}(a) also presents the inequivalent carbon atoms (C1-C6) in its unit cell. It is noted that the length of C1-C2 covalent bond is about 1.52 {\AA}, which is longer than that of C2-C3 (1.44 {\AA}), C3-C4 (1.41 {\AA}), C4-C5 (1.43 {\AA}), and C5-C6 (1.40 {\AA}) from our first principle calculations. Moreover, based on our band structure calculations as shown in Fig. \ref{fig2}(b) and \ref{fig2}(c), the valence and conduction energy bands meet at the Fermi level and show the features of a distorted Dirac cone in its first BZ [elaborated in Fig. \ref{fig2}(b)]. The Fermi velocity ($\upsilon$$_f$) of phagraphene in both \emph{k}$_\emph{x}$ and \emph{k}$_\emph{y}$ directions were obtained by fitting these two bands at \emph{k}$_\emph{i}$=\emph{K}$_\emph{i}$+\emph{q} (\emph{i} = \emph{x}, \emph{y}) to the expression of  \emph{f} = \emph{E}(\emph{q})/$\hbar$$|\emph{q}|$, which can be evaluated via the slope of the bands. In the \emph{k}$_\emph{x}$ direction, $\partial$\emph{E}/$\partial$\emph{k}$_\emph{x}$ =¡À26.8 eV{\AA} ($\upsilon$$_\emph{fx}$=6.48$\times$10$^5$m/s); while in the \emph{k}$_\emph{y}$ direction, the slope of the bands equals -25.8 eV{\AA} ($\upsilon$$_\emph{fy}$=6.24¡Á$\times$10$^5$m/s) and 14.2 eV{\AA} ($\upsilon$$_\emph{fy}$=3.43¡Á$\times$10$^5$m/s). These Fermi velocities are comparable with those of other carbon Dirac allotropes predicted in previous works, as listed in Table \ref{tab1}.

\begin{table}
\caption{\label{tab1}
Calculated total energy (\emph{E}$_t$)(eV/atom), planar carbon density (\emph{D})(atom/{\AA}$^2$), plane group (\emph{G}) and Fermi velocities ($\upsilon$$_f$)($\times$10$^5$m/s) of the most of carbon allotropes with Dirac cones from GGA-PBE results.}
\begin{tabular}{lccccccc}
\hline\hline
            & \emph{E}$_t$        & \emph{D}    & \emph{G}      & $\upsilon$$_f$ \\
\hline
$\alpha$-graphyne & -8.30  & 0.19 & \emph{P6m} & 6.77\\
$\beta$-graphyne  & -8.38  & 0.23 & \emph{P6m} & 3.87/4.35/6.77\\
$\delta$-graphyne & -8.49  & 0.26 & \emph{P6m} & 6.96\\
6,6,12-graphyne & -8.51  & 0.27 & \emph{Pmm} & 5.56/6.04/6.29(Cone I)\\
                &        &      &            & 1.69/2.18(Cone II)\\
phagraphene     & -9.03  & 0.37 & \emph{Pmg} & 3.43/6.24/6.48\\
graphene        & -9.23  & 0.38 & \emph{P6m} & 8.22\\
\hline\hline
\end{tabular}
\end{table}

In order to explore the origin of the distorted Dirac cone, partial charge density distributions of the two Dirac bands in proximity of the Fermi level were first investigated, as shown in Fig. \ref{fig2}(d). It is clear that the occupied states are inverted at the two sides of the Dirac point, which means the energy band inversion. Band-I and band-II, which are mainly contributed by the coupling of \emph{p$_z$} orbitals along \emph{a}$_1$ and \emph{a}$_2$ directions, cross each other and produce the Dirac point. In view of the coupling between \emph{p$_z$} orbitals leading to the formation of $\pi$-conjugated framework, we adopted a simple tight-binding (TB) Hamiltonian of the $\pi$-electrons to describe the electronic band structures in proximity of the Fermi level:

\begin{equation} \label{eq1}
H=-\sum_{ij}t_{ij}c_{i}^{+}c_{j}+h.c.
\end{equation}
where \emph{t}$_{ij}$ is the hopping energy of an electron between the \emph{i}-th and \emph{j}-th atoms, \emph{c}$_{i}$$^{+}$and \emph{c}$_{j}$ are the creation and annihilation operators, respectively. The distance-dependent hoping energy is determined using the formula: $t_{ij}=t_{0} \times exp[q \times (1-d_{ij}/d_{0})]$, with \emph{t}$_0$ = 2.70 eV, \emph{q} = 2.20, and \emph{d}$_0$ = 1.5 {\AA}. Through diagonalization of a 20$\times$20 matrix in the reciprocal space, the electronic band structures can be obtained, as indicated by the red dashed lines in Fig. \ref{fig2}(c). Obviously, this simple TB Hamiltonian in Eq. (\ref{eq1}) can reproduce the band structure of phagraphene in the vicinity of the Fermi level given by DFT calculations.

\begin{figure}
\begin{center}
\includegraphics[width=0.825\columnwidth]{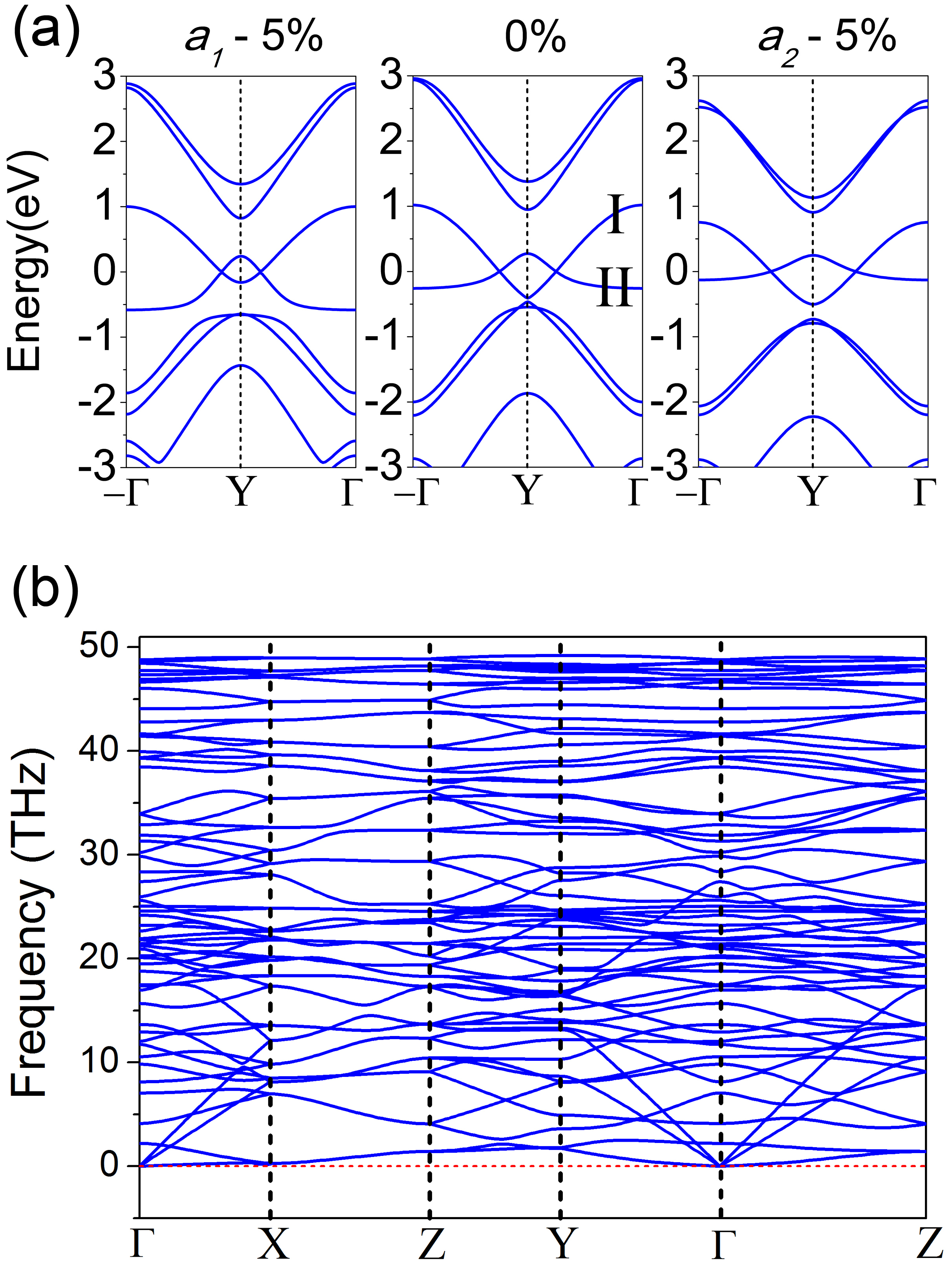}
\end{center}
\caption{\label{fig3}
(color online) (a) Band structures of phagrahene along $-\Gamma-Y-\Gamma$ line, with and without 5\% strain along \emph{a}$_1$ and \emph{a}$_2$ directions respectively. Fermi level has been set to zero; (b) Phonon dispersion of phagraphene.}
\end{figure}

As a matter of fact, most Dirac cones are indeed from the energy band crossing, which can be explained by a simple model: considering a 2-band system, the Hamiltonian can be written as:
\begin{equation}\label{eq2}
H(k)=\begin{bmatrix}
H_{11}(k)-E & H_{12}(k)\\
 H_{21}(k) & H_{22}(k)-E
\end{bmatrix}
\end{equation}
The appearance of Dirac cones corresponds to degenerate solutions of wave function of this Hamiltonian. In other word, the determinant of \emph{H}(\emph{k}) should be equal to zero and the following equations need to be satisfied:
\begin{equation}\label{eq3}
H_{11}(k)=H_{22}(k);H_{12}(k)=H_{21}(k)=0
\end{equation}
It is intrinsic that these three conditions in Eq. (\ref{eq3}) must be simultaneously fulfilled to have a degeneracy, which is known as the von Neumann-Wigner theorem \cite{R31,R32}. In most cases, $H_{11}(k)=H_{22}(k)$ can be fulfilled by the space-time inversion symmetries, but the 2D system symmetries cannot guarantee the existence of Dirac cones, because the number of variables (\emph{k}$_\emph{x}$ and \emph{k}$_\emph{y}$) is usually less than the number of equations to determine the Dirac points, which gives the reason of the rarity of 2D Dirac materials. As for phagraphene here, band-I and band-II in Fig. \ref{fig2}(d) or Fig. \ref{fig3}(a) can be simply considered as the two bands in above model, only they can appear in the vicinity of the Fermi level and invert to each other, the distorted Dirac cones are then produced at both sides of the space-symmetric and time-invariant \emph{k}-point Y. Although the determinant of its real Hamiltonian is hard to derive out analytical solutions, our numerical DFT and TB calculations all reveal that phagraphene happens to fulfill these three conditions simultaneously and presents distorted Dirac cones.

Its direction-dependent character can be further investigated by applying external strain, as shown in Fig. \ref{fig3}(a). The cone is stable under 5\% strain along \emph{a}$_1$ and \emph{a}$_2$ directions, respectively. Due to the band symmetry, both position and anisotropy of the Dirac cone can be tuned by external strain. For 5\% strain along \emph{a}$_1$ direction, the Dirac point moves from (0, 0.377, 0) to (0, 0.417, 0). The corresponding Fermi velocities change from ¡À26.8 eV{\AA} to ¡À23.7 eV{\AA} in the \emph{k}$_\emph{x}$ direction, from -14.2/25.8 eV{\AA} to -19.4/28.4 eV{\AA} in the \emph{k}$_\emph{y}$ direction. Things become different when a 5\% strain is added along \emph{a}$_2$ direction. Dirac point shifts to (0, 0.313, 0), and the corresponding Fermi velocities become ¡À28.1 eV{\AA} in the \emph{k}$_\emph{x}$ direction and -25.4/6.7 eV{\AA} in the \emph{k}$_\emph{y}$ direction.

In order to examine its dynamic stability, the phonon spectrum was calculated, as shown in Fig. \ref{fig3}(b), no imaginary frequency modes were observed. To further examine its thermal stability, a 3$\times$3 supercell was built to perform \emph{ab initio} molecular dynamics simulations. After being heated at room temperature (300 K) and 1000 K for 3 \emph{ps} with a time step of 1 \emph{fs}, no structural changes occur, except some fluctuations in the 2D plane. All these indicate that phagraphene is a new low-energy carbon allotrope with distorted Dirac cones. Considering 5- and 7-carbon ring dislocations and corresponding line defects have been observed in graphene \cite{R33}, phagraphene may be realized experimentally in near future.

In summary, systematic \emph{ab initio} evolutionary structure searches for 2D carbon networks identified a new low-energy 2D form of carbon, which is composed of 5-6-7 carbon rings and named as phagraphene. Thanks to its \emph{sp}$^2$-hybridization and density of atomic packing comparable to graphene, this 2D carbon structure with \emph{Pmg} plane group is lower in energy than most of the predicted 2D carbon allotropes. Both DFT and TB model confirm its distorted Dirac cone in the first BZ. The direction-dependent Dirac cones are further proved to be robust against the external strain with tunable Fermi velocities. This exotic structure is not only promising for fully investigating the massless Dirac fermions in 2D carbon electronic systems, but also helpful for understanding the topological and correlated phases in the corresponding photonic artificial lattices.

This work was jointly supported by the China Scholarship Council (No. 201408320093) and research projects (Grants No. BK20130859, No. 13KJB510019 and No. NY213010); X. F. Z. thanks the National Science Foundation of China (Grant No. 11174152), the National 973 Program of China (Grant No. 2012CB921900), the Program for New Century Excellent Talents in University (Grant No. NCET-12-0278), and the Fundamental Research Funds for the Central Universities (Grant No. 65121009); M. Z. thanks the support from the National Basic Research Program of China (Grant No. 2012CB932302) and the National Natural Science Foundation of China (Grant No. 91221101); A. R. O. thanks the National Science Foundation (EAR-1114313, DMR-1231586), DARPA (Grants No. W31P4Q1210008 and No. W31P4Q1310005), the Government of Russian Federation (No. 14.A12.31.0003), the Foreign Talents Introduction and Academic Exchange Program (No. B08040) and the support from SUNY 4E NoE.
Calculations were mainly performed on the cluster (INSPUR) of Peter Gr\"{u}nberg Reseach Center at Nanjing University of Posts and Communications and the cluster (QSH) in Oganov's lab at Stony Brook University.




\begin{references}

\bibitem{R01} A. H. Castro Neto, F. Guinea, N. M. R. Peres, K. S. Novoselov, and A. K. Geim, Rev. Mod. Phys. \textbf{81}, 109 (2009).

\bibitem{R02} K. I. Bolotin, K. J. Sikes, Z. Jiang, M. Klima, G. Fudenberg, J. Hone, P. Kim, and H. L. Stormer, Solid State Comm. \textbf{146}, 351 (2008).

\bibitem{R03} Y. Zhang, Y.-W. Tan, H. L. Stormer, and P. Kim, Nature \textbf{438}, 201 (2005).

\bibitem{R04} K. I. Bolotin, F. Ghahari, M. D. Shulman, H. L. Stormer, and P. Kim, Nature \textbf{462}, 196 (2009).

\bibitem{R05} K. S. Novoselov, A. K. Geim, S. V. Morozov, D. Jiang, Y. Zhang, S. V. Dubonos, I. V. Grigorieva, and A. A. Firsov, Science \textbf{306}, 666 (2004).

\bibitem{R06} J. Wang, S. Deng, Z. Liu, and Z. Liu, Natl. Sci. Rev. \textbf{2}(1), 22-39 (2015).

\bibitem{R07} S. Cahangirov, M. Topsakal, E. Akt\"{u}rk, H. \c{S}ahin, and S. Ciraci, Phys. Rev. Lett. \textbf{102}, 236804 (2009).

\bibitem{R08} H. Zhou, M. Zhao, X. Zhang, W. Dong, X. Wang, H. Bu, and A. Wang, J. Phys. Cond. Mat. \textbf{25}, 395501 (2013).

\bibitem{R09} D. Malko, C. Neiss, F. Vi\~{n}es, and A. G\"{o}rling, Phys. Rev. Lett. \textbf{108}, 086804 (2012).

\bibitem{R10} M. Zhao, W. Dong, and A. Wang, Sci. Rep. \textbf{3}, 3532 (2013).

\bibitem{R11} X.-F. Zhou, X. Dong, A. R. Oganov, Q. Zhu, Y. Tian, and H.-T. Wang, Phys. Rev. Lett. \textbf{112}, 085502 (2014).

\bibitem{R12} L. Z. Zhang, Z. F. Wang, S. X. Du, H.-J. Gao, and Feng Liu, Phys. Rev. B. \textbf{90}, 161402(R) (2014).

\bibitem{R13} W. Li, M. Guo, G. Zhang, and Y.-W. Zhang, Phys. Rev. B \textbf{89}, 205402 (2014).

\bibitem{R14} K. S. Novoselov, A. K. Geim, S. V. Morozov, D. Jiang, M. I. Katsnelson, I. V. Grigorieva, S. V. Dubonos, and A. A. Firsov, Nature \textbf{438}, 197 (2005).

\bibitem{R15} R. H. Baughman, H. Eckhardt, and M. Kertesz, J. Chem. Phys. \textbf{87}, 6687 (1987).

\bibitem{R16} A. N. Enyashin and A. L. Ivanovskii, Phys. Status Solidi B \textbf{248}, 1879 (2011).

\bibitem{R17} N. Narita, S. Nagai, S. Suzuki, and K. Nakao, Phys. Rev. B \textbf{58}, 11009 (1998).

\bibitem{R18} H. Huang, W. Duan, and Z. Liu, New J. Phys. \textbf{15}, 023004 (2013).

\bibitem{R19} Y. Liu, G. Wang, Q. Huang, L. Guo, and X. Chen, Phys. Rev. Lett. \textbf{108}, 225505 (2012).

\bibitem{R20} H. Huang, Y. Li, Z. Liu, J. Wu, and W. Duan, Phys. Rev. Lett. \textbf{110}, 029603 (2013).

\bibitem{R21} L.-C. Xu, R.-Z. Wang, M.-S. Miao, X.-L. Wei, Y.-P. Chen, H. Yan, W.-M. Lau, L.-M. Liu, and Y.-M. Ma, Nanoscale \textbf{6}, 1113 (2014).

\bibitem{R22} A. R. Oganov and C. W. Glass, J. Chem. Phys. \textbf{124}, 244704 (2006).

\bibitem{R23} C. W. Glass, A. R. Oganov, and N. Hansen, Comput. Phys. Commun. \textbf{175}, 713 (2006).

\bibitem{R24} Q. Zhu, L. Li, A. R. Oganov, and P. B. Allen, Phys. Rev. B \textbf{87}, 195317 (2013).

\bibitem{R25} P. E. Bl\"{o}chl, Phys. Rev. B \textbf{50}, 17953 (1994).

\bibitem{R26} G. Kresse and J. Furthmuller, Phys. Rev. B \textbf{54}, 11169 (1996).

\bibitem{R27} G. Kresse and J. Furthmuller, Comput. Mater. Sci. \textbf{6}, 15 (1996).

\bibitem{R28} J. P. Perdew, K. Burke, and M. Ernzerhof, Phys. Rev. Lett. \textbf{77}, 3865 (1996).

\bibitem{R29} A. Togo, F. Oba, and I. Tanaka, Phys. Rev. B \textbf{78}, 134106 (2008).

\bibitem{R30} Serial number in Fig. \ref{fig1}.(b). Name, (Atomic number of the primitive cell), [Reference(s)]: 1. Graphene, (\textbf{2}), \cite{R05}; 2. C$31$-sheet, (\textbf{4}), \cite{R34}; 3. Biphenylene, (\textbf{6}), \cite{R35,R36}; 4. $\alpha$-graphyne, (\textbf{8}), \cite{R09}; 5. Planar T-graphene, (\textbf{8}), \cite{R19}; 6. Pentaheptite, (\textbf{8}), \cite{R37}; 7. S-graphene, (\textbf{8}), \cite{R21}; 8. Rectangular heackelite, (\textbf{8}), \cite{R38}; 9. Pentahexoctite, (\textbf{8}), \cite{R39}; 10. Net-W, (\textbf{8}), \cite{R36}; 11. HOP Graphene, (\textbf{10}), \cite{R40}; 12. C$65$-sheet, (\textbf{10}), \cite{R34}; 13. Pza-C$10$, (\textbf{10}), \cite{R41}; 14. $\gamma$-graphyne, (\textbf{12}), \cite{R17}; 15. OPG-L, (\textbf{12}), \cite{R42}; 16. OPG-Z (\textbf{12}), \cite{R42}; 17. C$41$-sheet, (\textbf{12}), \cite{R34}; 18. C$63$-sheet, (\textbf{12}), \cite{R34}; 19. C$64$-sheet or BPC, (\textbf{12}), \cite{R34,R43}; 20. Structrure-$9$ or 14,14,14-graphyne, (\textbf{12}), \cite{R16,R18}; 21. Oblique heackelite, (\textbf{12}), \cite{R38}; 22. Fused pentagon network, (\textbf{14}), \cite{R44}; 23. Hexagonal heackelite, (\textbf{16}), \cite{R38}; 24. 6,6,12-graphyne, (\textbf{18}), \cite{R09}; 25. $\beta$-graphyne, (\textbf{18}), \cite{R09}; 26. $\delta$-graphyne, (\textbf{20}), \cite{R10}; 27. Phagraphene, (\textbf{20}). All above structures have been presented in our supporting POSCARS file, see supplemental material at

\bibitem{R31} J.V. Neumann and E. Wigner, Physik Z. \textbf{30}, 467 (1929).

\bibitem{R32} L.D. Landau and L.M. Lifshitz, Quantum Mechanics Non-Relativistic theory 3rd ed., (Butterworth-Heinemann, Stoneham, MA,1981), Sec. 79.

\bibitem{R33} O. V. Yazyev and Y. P. Chen, Nat. Nano. \textbf{9}, 755 (2014).


\bibitem{R34} H. Lu and S.-D. Li, J. Mater. Chem. C \textbf{1}, 3677 (2013).

\bibitem{R35} M. A. Hudspeth, B. W. Whitman, V. Barone, and J. E. Peralta, ACS Nano \textbf{4}, 4565 (2010).

\bibitem{R36} X.-Q. Wang, H.-D. Li, and J.-T. Wang, Phys. Chem. Chem. Phys. \textbf{15}, 2024 (2013).

\bibitem{R37} V. H. Crespi, L. X. Benedict, M. L. Cohen, and S. G. Louie, Phys. Rev. B \textbf{53}, R13303 (1996).

\bibitem{R38} H. Terrones, M. Terrones, E. Hern¨¢ndez, N. Grobert, J. C. Charlier, and P. M. Ajayan, Phys. Rev. Lett. \textbf{84}, 1716 (2000).

\bibitem{R39} B. R. Sharma, A. Manjanath, and A. K. Singh, Sci. Rep. \textbf{4}, 7164 (2014).

\bibitem{R40} B. Mandal, S. Sarkar, A. Pramanik, and P. Sarkar, Phys. Chem. Chem. Phys. \textbf{15}, 21001 (2013).

\bibitem{R41} X. Luo, L. Liu, Z. Hu, W. Wang, W. Song, F. Li, S. Zhao, H. Liu, H.-T. Wang, and Y. Tian, J. Phys. Chem. Lett. \textbf{3}, 3373 (2012).

\bibitem{R42} C. Su, H. Jiang, and J. Feng, Phys. Rev. B \textbf{87}, 075453 (2013).

\bibitem{R43} G. Brunetto, P. A. S. Autreto, L. D. Machado, B. I. Santos, R. P. B. dos Santos, and D. S. Galv\~{a}o, J. Phys. Chem. C \textbf{116}, 12810 (2012).

\bibitem{R44} M. Mina and O. Susumu, Appl. Phys. Express \textbf{6}, 095101 (2013).





\end{references}
\end{document}